\documentclass{iaus}
\usepackage{graphicx}

\title[Multiplicity Study of exoplanet host stars] %% give here short title %%
{Multiplicity Study of exoplanet host stars: \newline The HD\,3651\,AB system}
\author[Mugrauer et al.]   %% give here short author list %%
{Mugrauer M.$^1$ \thanks{markus@astro.uni-jena.de}, Seifahrt A.$^2$, Neuh\"{a}user R.$^1$, Mazeh
T.$^3$, and Schmidt T.$^1$}

\affiliation{$^{1}$Astrophysikalisches Institut, Universit\"at Jena, Schillerg\"a{\ss}chen 2-3, 07745 Jena, Germany\\
$^{2}$European Southern Observatory, Karl-Schwarzschild-Str. 2, 85748 Garching, Germany\\
$^{3}$Tel Aviv University, Tel Aviv 69978, Israel\\ [\affilskip]}

\begin{document}

\maketitle

\begin{abstract}
We report on new results of our multiplicity study and present our follow-up observations of the
recently detected faint companion HD\,3651\,B, the first directly imaged T dwarf companion of an
exoplanet host star. \keywords{(stars:) binaries: visual, stars: low-mass, brown dwarfs}
\end{abstract}

\firstsection % if your document starts with a section,
\section{Introduction}
In the course of our observations to search for wide faint companions to planet host stars, carried
out in the infrared (H-band) with SofI at the NTT and UFTI at the UKIRT, we could already find
several new companions (Mugrauer et al. 2004a \& b, 2005, 2006b). Most of these objects are
low-mass stars with masses which range from the substellar-stellar mass border at about
0.075\,M$_{\odot}$ up to about 0.5\,\,M$_{\odot}$.

Two companions turned out to be white dwarfs, whose true nature could be revealed with
spectrophotometric measurements. Gl\,86\,B is the first known close white dwarf companion of an
exoplanet host star, separated from its primary by only 20\,AU (Mugrauer \& Neuh\"{a}user 2005 and
Mugrauer et al. 2006a). Its white dwarf nature could also be confirmed by combining radial velocity
data with NACO astrometry, obtained during the last years (Lagrange et al. 2006). These
observations will finally constrain the orbital parameters of the companion and also allow to
determine its mass. A first estimate shows that the white dwarf companion (m$\sim$0.5\,M$_{\odot}$)
revolves around the planet host star Gl\,86\,A on an eccentric $e=0.4$ orbit with a semi-major axis
of only 18.4\,AU. If we assume that the system was about 1.4 times closer before the mass loss of
the white dwarf progenitor star (Weidemann 2000), such a close companion yields only a small long
time stable region for planets (only 1.7\,AU, Holman \& Wiegert 1997). The same holds for the size
of the protoplanetary disk in which the planet formation process took place (Pichardo et al. 2005),
posing a problem to the theory of planet formation. Nevertheless, a massive planet
($msin(i)=$4\,M$_{\rm Jup}$, Queloz et al. 2000) had been formed in this environment and finally
also survived the post main sequence evolution of the white dwarf progenitor star.

A further white dwarf companion could be detected around the planet host star HD\,27442\,A. The
white dwarf nature of HD\,27442\,B (WDS binary, Worley \& Douglas 1997) could be revealed with SofI
photometry and ISAAC and FORS spectroscopy (Mugrauer et al. 2006, submitted to A\&A). This
companion is separated from the planet host star by already $\sim$240\,AU, hence this is a less
challenging environment for planet formation than the Gl\,86\,AB system. If planets can form in a
Gl\,86\,AB like system and also survive the post main sequence evolution of the white dwarf
progenitor star, planets should also exist in wider such evolved stellar systems with white dwarf
companions. However, the HD\,27442\,AB system is a special case because of the nature of the planet
host star which is an already evolved subgiant (Butler et al. 2001). With its white dwarf companion
and the subgiant planet host star, the HD\,27442\,AB system is the most evolved planet hosting
stellar system, presently known.

\section{The HD\,3651\,AB system}

Recently, we reported the detection of a new very faint companion 43\,arcsec north-west of the
planet host star HD\,3651\,A (Mugrauer et al. 2006c). This companion, HD\,3651\,B, was followed by
us in four observing epochs with UFTI at UKIRT and SofI at NTT, clearly confirming that it shares
the proper motion of the nearby planet host star. The averaged SofI and UFTI H-band photometry of
the companion is H=16.75$\pm$0.16\,mag, i.e. this is the faintest co-moving companion of an
exoplanet host star, presently known. HD\,3651\,B is not detected in the visible (POSS Plates) but
appears only as a faint source in the infrared, i.e. it is a cool brown dwarf companion, the first
of its kind directly imaged as companion of an exoplanet host star. The H-band photometry is
consistent with a T dwarf of spectral type T7 to T8 (800 - 900\,K). HD\,3651\,B is one of the
faintest T-dwarfs ever imaged.

During the astrometrical analysis process of our imaging data we also studied the 2MASS J-, H-, and
K$_{\rm S}$-band images (Fig.\,1) in order to check whether the companion is detected in the 2MASS
data, providing additional astrometric and photometric data. Indeed, there is a very faint noisy
source ($S/N<5$) which can be found in the 2MASS J- and H-band images exactly at the expected
position of HD\,3651\,B. In the K$_{\rm S}$-band there is a source detected slightly westward from
the expected position. This faint source is not listed in the 2MASS Point Source Catalog (Skrutskie
et al. 2006). It might be the companion, but could also be spurious. Due to the faintness of this
source, and because there are several comparable bright speckles detected in the 2MASS images, we
did not use the 2MASS data for astrometric or photometric analysis. In contrast, in our four images
obtained with UFTI and SofI HD\,3651\,B is clearly detected, hence accurate astrometric and
photometric measurements are possible.

\begin{figure}[htb]
\resizebox{\hsize}{!}{\includegraphics[width=12cm]{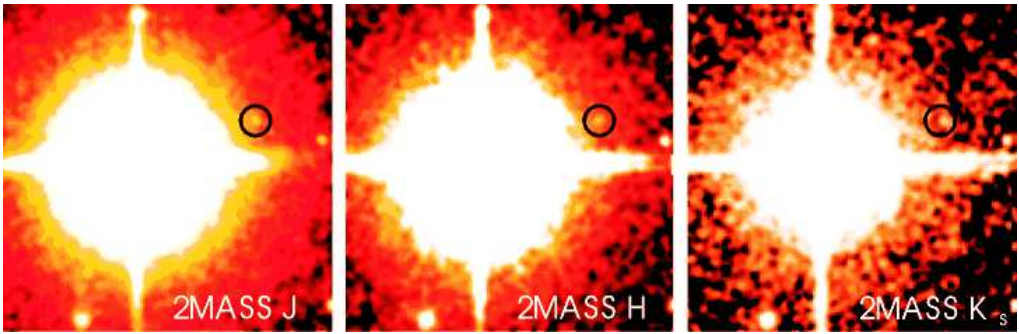}}\caption{The 2MASS J-, H-, and
K$_{\rm S}$-band Atlas images. The expected position of HD\,3651\,B is at the center of the black
circles. We detected a faint source in 2MASS J-, and H-band images, which is not visible in the
2MASS K$_{\rm S}$-band image. This might be the co-moving companion HD\,3651\,B, but could be
spurious.} \label{2MASS}
\end{figure}

\begin{figure}[htb]
\resizebox{\hsize}{!}{\includegraphics[height=5cm]{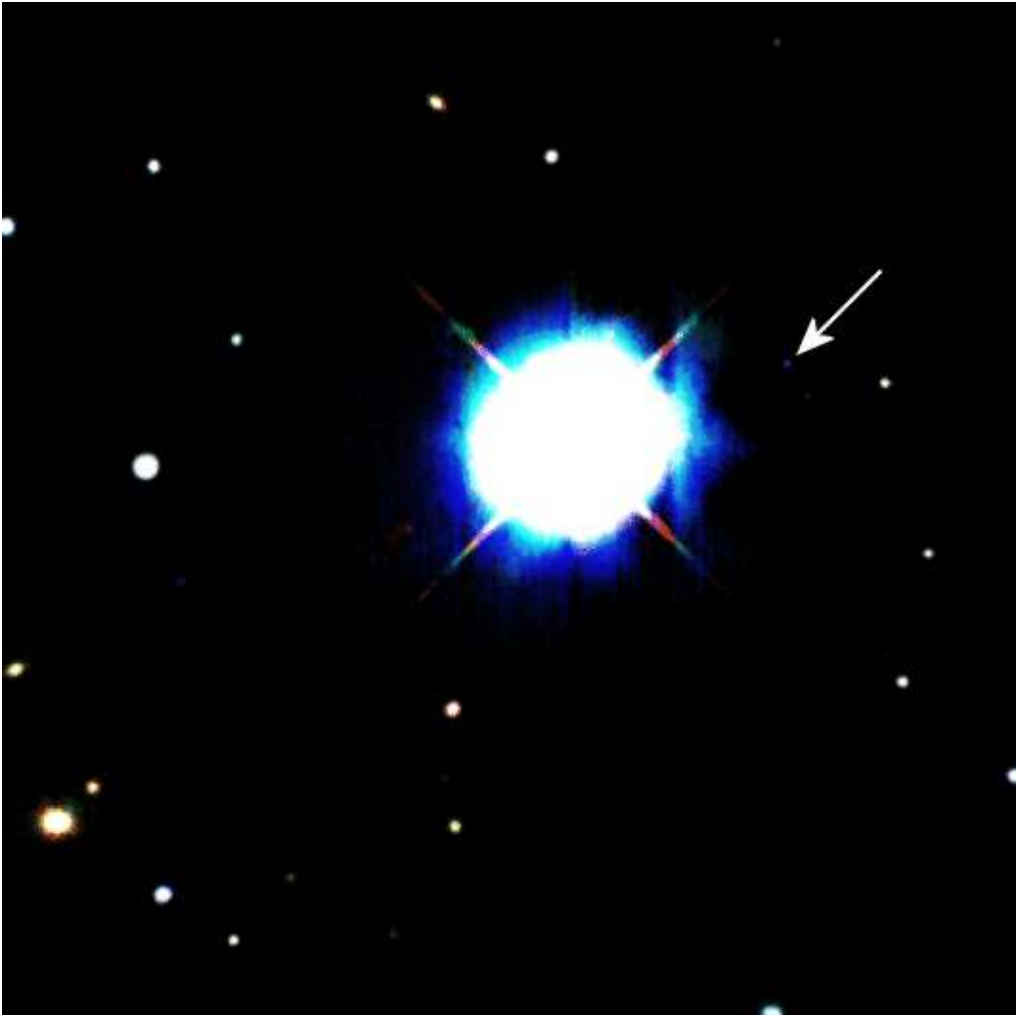}}\caption{This is a J- (blue), H-
(green), and K-band (red) color composite-image of the exoplanet host star HD\,3651\,A (bright star
in the center) taken with the infrared camera OMEGA-CASS at the 3.5\,m telescope at Calar Alto
observatory. The infrared imaging is carried out in the high resolution mode (0.2\,arcsec/pixel).
The detected faint co-moving companion HD\,3651\,B is marked with an arrow.}\label{cahapic}
\end{figure}

In addition to our UFTI and SofI imaging we obtained follow-up J-, H-, and K-band imaging data of
the new companion with OMEGA-CASS at the 3.5\,m telescope of the Calar Alto observatory (south
Spain). The companion was observed 9.8\,min in J- and H-, and 35.5\,min in K-band. The J- and
H-band OMEGA-CASS images are both the average of 10 frames, each the average of 49
1.2\,s-integrations. The K-band image is the average of 29 frames, each the average of 49
1.5\,s-integrations. We measure the infrared photometry of HD\,3651\,B in all OMEGA-CASS images and
obtain: J$=15.98$$\pm$0.12 \,mag, H$=16.80$$\pm$0.08\,mag, and K$=16.53$$\pm$0.13\,mag (see
Fig.\,2). The companion can be plotted then in a color-magnitude diagram (see Fig.\,3) together
with comparison brown dwarfs from Knapp et al (2004). The OMEGA-CASS photometry confirms our
conclusion that the companion is a late T dwarf. Its photometry is fully consistent with late T
dwarfs like Gl\,570\,D (Burgasser et al. 2000) or 2MASSJ0415 (Burgasser et al. 2002). Our spectral
type estimate was most recently confirmed by Luhman et al. (2006) and Burgasser (2006).

\begin{figure}[htb]
\resizebox{\hsize}{!}{\includegraphics[height=5cm]{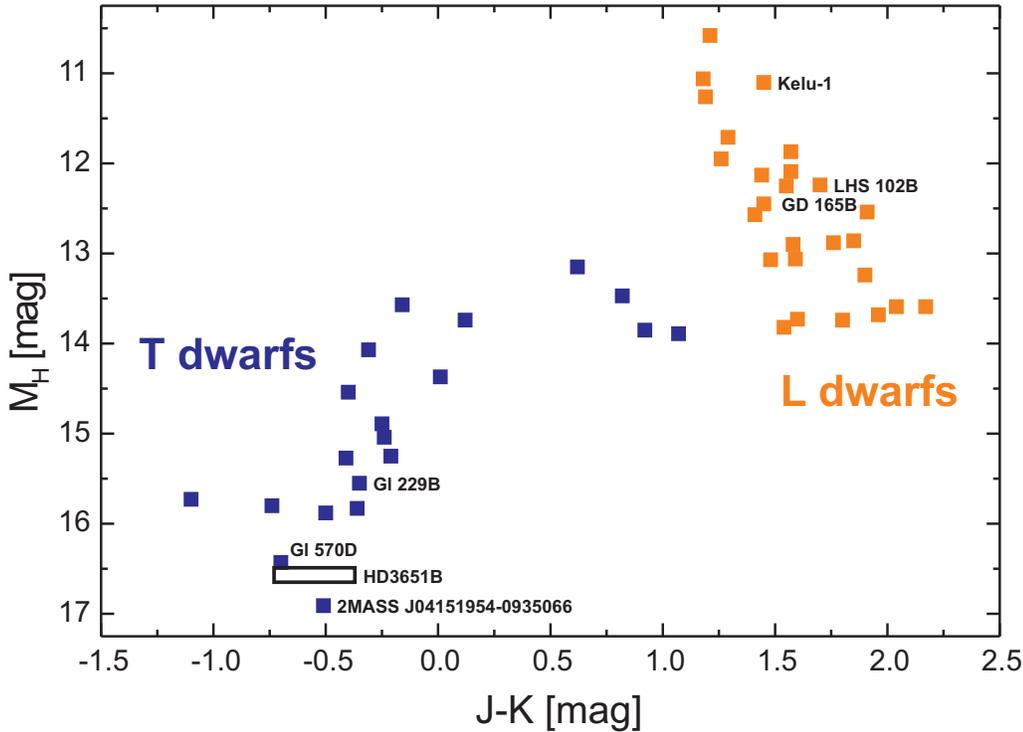}} \caption{HD\,3651\,B in the
(J$-$K)$-$M$_{\rm H}$ diagram together with comparison brown dwarfs from Knapp et al. (2004).
HD\,3651\,B is one of the faintest T dwarfs presently known.}\label{magcol}
\end{figure}

\end{document}